\documentclass{elsart}
\usepackage[OT4]{fontenc}
\usepackage[utf8]{inputenc}
\usepackage{amsfonts}
\usepackage{amssymb}
\usepackage{graphicx}

\begin{document}
\begin{frontmatter}

\title{Correlations in commodity markets}
\author{Paweł Sieczka}, \ead{psieczka@if.pw.edu.pl} \author{Janusz A. Hołyst \corauthref{jh}} \ead{jholyst@if.pw.edu.pl}

\corauth[jh]{corresponding author}

\address{Faculty of Physics, Center of Excellence for Complex Systems Research, Warsaw University of Technology, Koszykowa 75, PL-00-662 Warsaw, Poland}

\date{\today}

\begin {abstract}
In this paper we analyzed dependencies in commodity markets investigating correlations of future contracts for commodities over the period 1998.09.01 - 2007.12.14. We constructed a minimal spanning tree based on the correlation matrix. The tree provides evidence for sector clusterization of investigated contracts. We also studied dynamical properties of commodity dependencies. It turned out that the market was constantly getting more correlated within the investigated period, although the increase of correlation was distributed nonuniformly among all contracts, and depended on contracts branches.
\end{abstract}


\end{frontmatter}

\section{Introduction}
Commodity markets are in their origins the prime and the most basic markets rooted in times when people were exchanging goods even before money was invented. Today's commodity markets are mature and highly developed institutions, playing a very important role in modern economy. They are not only places of goods exchange, but also a theater of speculative activity.

Nowadays, when we are used to highly so\-phi\-sti\-ca\-ted financial instruments, including cre\-dit de\-ri\-va\-ti\-ves, contracts for other contracts for some artificial underlying instruments,  etc., commodities seem to be rather old-fa\-shio\-ned. Yet, they remain important, not only due to their being primary raw materials for other stages of economic activity, but also because they can be a reliable measure of value, especially in times of crisis or other historical turbulence.

Commodities, traded at free markets, follow the rules of the efficient market hypothesis \cite{Fama}, the same as stocks, currencies, and others. Changes of their prices are, therefore, random and in major part unpredictable. A model which reflects this property is the geometric Brownian motion of prices, the core of the Black-Scholes theory \cite{Bouchaud}.  However, real prices of financial assets deviate from the Brownian behavior, what has been clearly shown by investigations using different tools of econophysics. The autocorrelation function of the absolute returns decays as a power law with an exponent $-0.3$ \cite{Gopikrishnan}. The returns are weakly correlated \cite{Bouchaud,Voit} and show persistent behavior of their sign \cite{Zhang,Sieczka}. All those observations for stock markets should be also in general  valid for commodity markets, despite observed differences such as a \textit{spatial arbitrage} effect \cite{Roehner}, or different multifractal properties \cite{Matia}. Matia \textit{et al}. \cite{Matia2} showed also that the prices of commodity futures obey a different scaling law from the prices of spots. The former are more similar to stocks in this aspect.

It is well known that stocks of different firms are mutually correlated in a way that cannot be totally explained by the random matrix theory \cite{Laloux,Plerou,Kwapien, Coelho}.  The correlation coefficients of stock price returns can be used to obtain a minimal spanning tree and  associated with it a hierarchical tree of the subdominant ultrametric space, which was done by Mantegna \cite{Mantegna}. He detected grouping of firms of a similar profile in the minimal spanning tree. This effect can be reproduced neither by the random model of uncorrelated time series, nor by the one-factor model \cite{Bonanno}.

In this paper we analyzed cross-correlations in commodity markets. We created the correlation matrix and corresponding correlation-based metric. Using the correlation metric we created a minimal spanning tree of investigated contracts, looking for sector clusterization. We also examined dynamic properties of correlations, finding out that commodity contracts were getting more and more correlated, and mean distances in a corresponding minimal spanning tree were smaller and smaller in the course of time. However, individual contracts contributed to the increase of mean correlation differently. Their idiosyncratic contribution to the correlations was characterized by an introduced quantity (strength) and its evolution.

The motivation for our research was a growing interest of investors and mass media in commodity markets. Skyrocketing oil prices were expected to reach the level of 200 USD per barrel during one week, and went down under 100 USD during another. We wanted to investigate the behavior of the commodity markets with the tools of complex system physics. We found the mirroring of the  specific situation of last years in a time dependent picture of commodity prices correlations.

\section{The data}

We investigated 35 future contracts for commodities tra\-ded at di\-ffe\-rent mar\-kets. Futures rather than spots were examined as the data were more accessible. 

 We used data from: Chicago Board of Trade (CBOT), Chicago Mercantile Exchange (CME), IntercontinentalExchange (ICE), Kansas City Board of Trade (KCBT), London Metal Exchange (LME), Minneapolis Grain Exchange (MGEX), New York Board of Trade (NYBOT), New York Mercantile Exchange (NYMEX), Winnipeg Commodity Exchange (WCE).   For today's investors in the globalized world market a contract traded in London or in Chicago is only another financial instrument that they can buy or sell no matter where. 

Table \ref{tab_list} presents the list of investigated contracts, their symbols, and  symbols of corresponding markets. All the contracts were quoted in USD. We took under consideration day closing prices.

\begin{table}
\begin{tabular}{ | c c c | c c c |}
\hline
symbol & name & market & symbol & name & market\\ 
\hline
AA\_F & aluminum alloy & LME   & MW.F &wheat spring  & MGEX    \\
 AL\_F& aluminum  & LME     & NG.F &natural gas  & NYMEX    \\
BO.F &soybean oil  & CBOT    & NI\_F &nickel  & LME    \\
C.F & corn & CBOT    &  OJ.F& orange juice  & NYBOT    \\
CC.F &cocoa  &NYBOT    &  PA.F& palladium  & NYMEX    \\
CL.F &crude oil  & NYMEX    & PB.F & pork bellies  & CME    \\
 CO\_F& copper & LME    & PL.F & platinum & NYMEX   \\
CT.F & cotton  & NYBOT    & RR.F & rough rice & CBOT   \\
FC.F &feeder cattle  &CME    & RS.F & canola   & WCE    \\
GC.F & gold  & NYMEX    & S.F &soybean  & CBOT    \\
 HG.F& copper  &NYMEX    & SB.F & sugar  & NYBOT   \\
HO.F & heating oil & NYMEX   & SC.F & brent oil  & ICE    \\
KC.F &coffee  &  NYBOT   & SI.F &  silver & NYMEX   \\
KW.F & wheat   & KCBT    & SM.F &soybean meal  & CBOT    \\
LB.F & lumber  & CME    & TI\_F & tin & LME    \\
 LC.F& live cattle  & CME    & W.F & wheat & CBOT \\
LE\_F & lead  & LME   & ZI\_F & zinc & LME \\
LH.F &  lean hogs & CME   &   &      &\\

\hline
\end{tabular} 
\caption{List of investigated future contracts for commodities in the alphabetical order of their symbol. }
\label{tab_list}
\end{table}

\section{Correlations}
Let $P_i(t)$ be a day closing price of a contract $i$ at time $t$. From logarithmic returns $r_i=\log(P_i(t+1))-\log(P_i(t))$ we calculated a Pearson correlation coefficient:
\begin{equation}
C_{ij}=\frac{\langle r_i r_j\rangle -\langle r_i\rangle\langle r_j\rangle}{\sqrt{(\langle r_i^2\rangle-\langle r_i\rangle^2)(\langle r_j^2\rangle-\langle r_j\rangle^2)}}.
\end{equation}

The correlation coefficients $C_{ij}$ were computed for all pairs of futures from Tab. \ref{tab_list} over a span between  1998.09.01 and 2007.12.14. The average $\langle ... \rangle$ was calculated for the whole period, but only for days when all the contracts were traded. There were $T=2190$ overlapping records in the mentioned period. 

\begin{figure}
\includegraphics[scale=0.6, angle=-90]{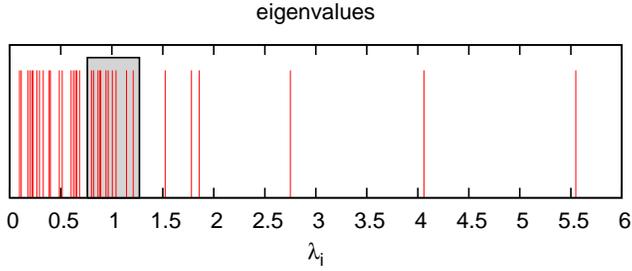} 
\caption{Eigenvalues of the correlation matrix. The gray re\-ctan\-gle co\-rres\-ponds to the area of a ran\-dom ma\-trix.}
\label{rys_eigenvalues}
\end{figure}

One could expect that due to long, compared to a number of assets, time series, the correlation matrix would have a low level of noise. According to the random matrix theory (RMT) \cite{Laloux}, the eigenvalues spectra of a correlation matrix for $N$ uncorrelated Gaussian time series of the length $T$  is bounded by a maximum $\lambda_{max}$ and a minimum $\lambda_{min}$ value, which is equal to:
\begin{equation}
\lambda_{min}^{max}=1+\frac{1}{Q}\pm2\sqrt{\frac{1}{Q}},
\end{equation}
where $Q=T/N$. In our case $Q\approx 62.57$ and $\lambda_{min}\approx 0.76$, $\lambda^{max}\approx 1.27$.  In figure \ref{rys_eigenvalues} the eigenvalue spectrum is presented. The majority of the eigenvalues lies outside the RMT region. 

\begin{figure}
\begin{center}
\includegraphics[scale=1]{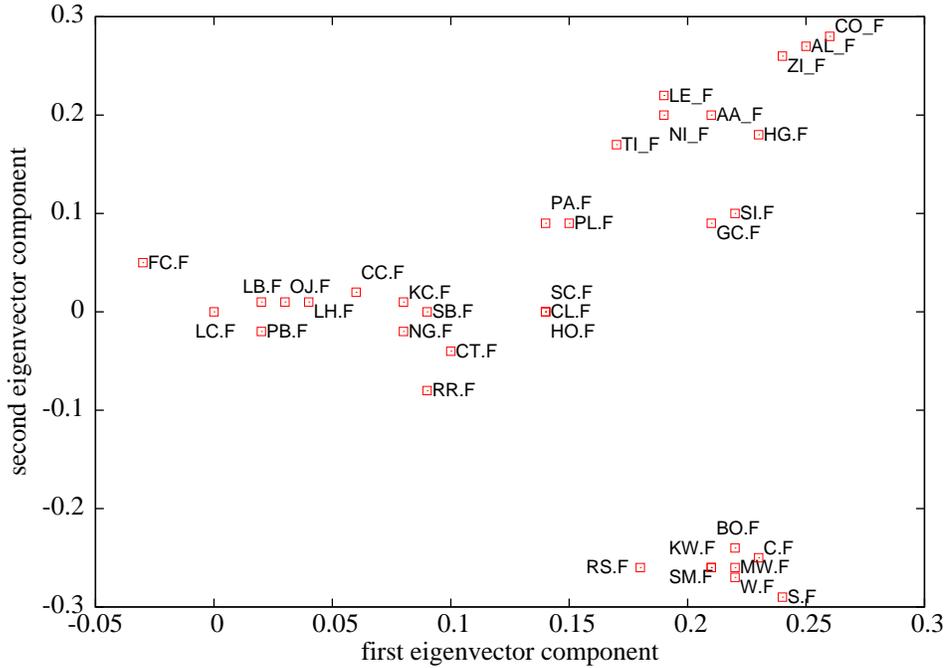} 
\caption{Two eigenvectors corresponding to the two largest eigenvalues of the correlation matrix. Each point can be identified as a contribution of a certain contract to an eigenvector. }                                                                               \label{rys_eigenvectors}                                                                      \end{center}

\end{figure}

Plotting components of two eigenvectors corresponding to the two largest eigenvalues shows a clustering structure of the correlation matrix \cite{Ausloos}. Assets of specific sectors depend on common factors in a similar way and give a similar contribution to the eigenvectors (fig. \ref{rys_eigenvectors}).

Following Mantegna \cite{Mantegna} we computed a metric distance matrix,

\begin{equation}
d_{ij}=\sqrt{2(1-C_{ij})}.
\end{equation}
The function $d_{ij}$ is a well-defined metric measure. It measures a distance between two time series, that is in our case between returns of two commodity futures. The closer they lie in the sense of the metric, the more correlated they are. 

The distance matrix $d_{ij}$ determines a weighted fully connected graph of correlation distances. Being symmetric, it has $N(N-1)$ independent elements, so does the correlation matrix $C_{ij}$. 

For a weighted network associated with $d_{ij}$ we can create a minimal spanning tree (MST). A spanning tree of a weighted graph $G$ is a tree that contains all vertices of $G$ and links of the tree are a subset of the links of $G$. A minimal spanning tree of a graph $G$ has the lowest sum of weights among all spanning trees of $G$. A MST of the distance matrix $d_{ij}$ has $N-1$ links. It pictures only the most important interactions, and hence is a useful tool for correlation visualization. 

\begin{figure}
\begin{center}
\includegraphics[ scale=0.35]{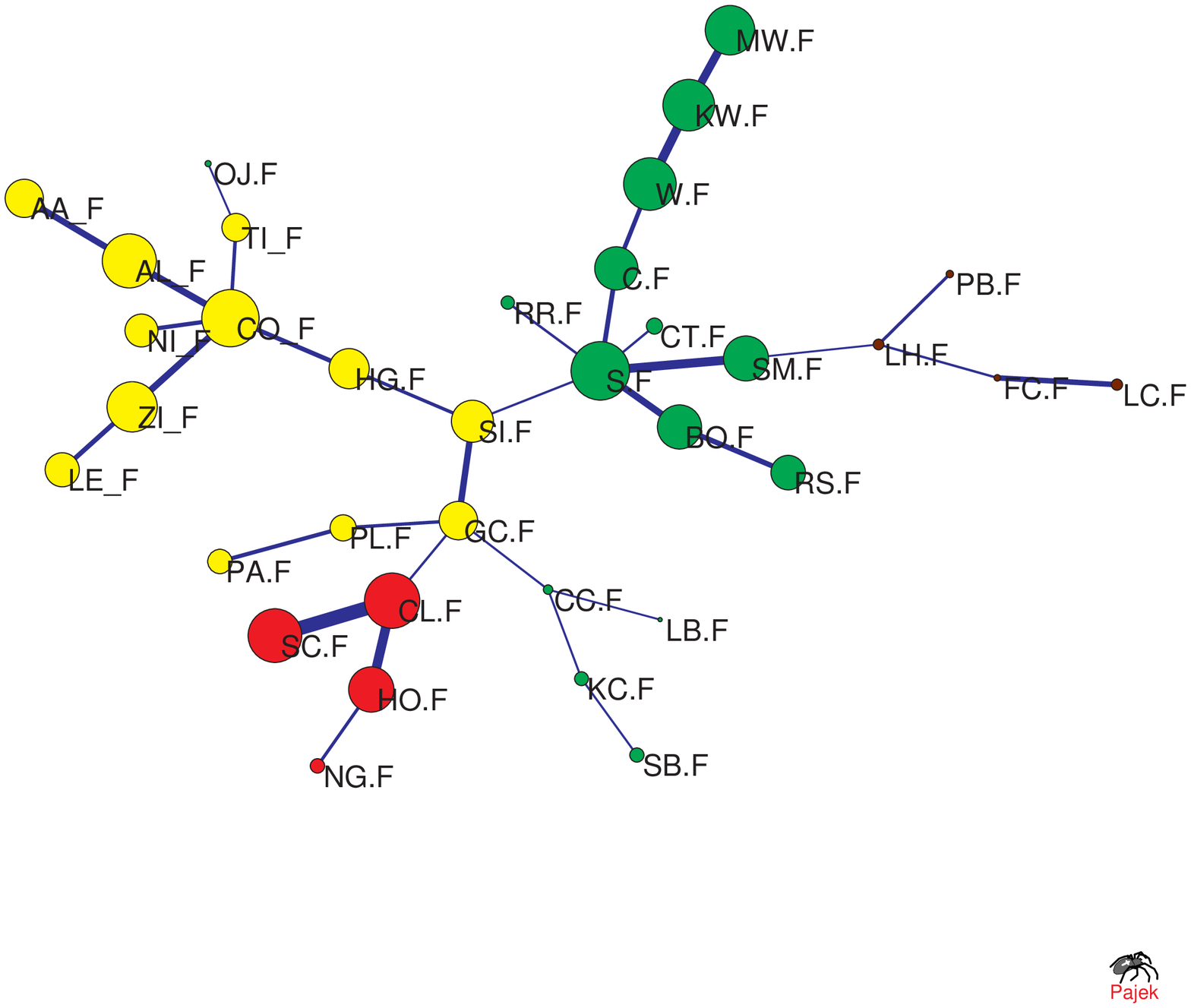}

\caption{(Color online) MST composed of 35 future contracts from tab. \ref{tab_list} with corresponding weights. For a given node its radius is proportional to its strength. Color denotes the branch: yellow -metals, red -fuels, green -plants, brown -animals. }                                                                       \end{center}
\label{rys_MST}
\end{figure}

 We calculated node strength defined as:
\begin{equation}
S_i=\sum_{j\neq i} \frac{1}{d_{ij}}. 
\label{sila}
\end{equation}
We also created a MST based on the metric distance  using Prim's algorithm \cite{Prim}.
Figure \ref{rys_MST} presents the MST with weights corresponding to distances between given nodes, and a node radius proportional to their strength. We used colors to distinguish different branches: metals with yellow, fuels with red, plant products with green, and animals with brown. The sector clusterization is clearly visible. We can identify connected subgraphs of specific profiles: metals (containing all listed metals), fuels (CL.F, SC.F, HO.F, NG.F), grains (S.F, C.F, W.F, KW.F, MW.F, RR.F, BO.F, RS.F, SM.F), animals (LB.F, PB.F, FC.F, LC.F), plant products (CC.F, KC.F, SB.F, LB.F) and outsiders (OJ.F, CT.F). One can also observe market clusterization, for example all metals from LME form a connected subgraph.

To find out the importance of a vertex in a graph different measures can be used. One of them is node strength (\ref{sila}), which informs how much  a given vertex is correlated with the others. Another is a node degree, defined as a number of links attached to a given node, or betweennes centrality, which is defined as \cite{Caldarelli}:
\begin{equation}
 B(i)=\sum_{(j, l)}\frac{\sigma_{jl}(i)}{\sigma_{jl}},
\end{equation}
where $\sigma_{jl}$ is the number of the shortest paths going from j to l and $\sigma_{jl}(i)$ is the number of the shortest paths from $j$ to $l$ passing through $i$. The sum is over indices $j$, $l$ fulfilling the condition: $i\neq j\neq l\neq i$.

For the minimal spanning tree (fig. \ref{rys_MST}), treated as an unweighted graph, we computed betweenness centrality of each node, its strength and degree (tab. \ref{tab:centrality}). The vertices with highest degrees are: SO.F 6, CO\_F 5, GC.F 4. The strongest are: S.F 27.7, CO\_F  27.62, CL.F 27.50, and the most between: SI.F 0.67, S.F 0.59, GC.F 0.48.

\begin{table}
\begin{tabular}{|c c c c | c c c c|}
\hline
symbol & deg. & strength & betw. & symbol & deg. & strength & betw.\\ 
\hline

AA\_F & 1 & 26.42 & 0   & MW.F & 1  & 27.13 & 0    \\
 AL\_F& 2  & 27.42 & 0.06     & NG.F & 1  & 24.91 & 0    \\
BO.F & 2  & 26.79 & 0.06    & NI\_F &1  & 26.08 & 0   \\
C.F & 2 & 26.73 & 0.16    &  OJ.F& 1  & 24.38 & 0    \\
CC.F & 3  & 24.59 & 0.17    &  PA.F& 1  & 25.54 & 0    \\
CL.F & 3  & 27.50 & 0.17    & PB.F & 1 & 24.46 & 0    \\
 CO\_F& 5 & 27.62 & 0.37    & PL.F & 2 & 25.65 & 0.06   \\
CT.F & 1  & 25.02 & 0    & RR.F & 1 & 24.84 & 0   \\
FC.F & 2  &24.41 & 0.06    & RS.F & 1   & 26.18 & 0   \\
GC.F & 4  & 26.43 & 0.48    & S.F & 6  &  27.70 & 0.59    \\
 HG.F& 2 &26.54  & 0.37    & SB.F & 1  & 24.90 & 0   \\
HO.F & 2 & 26.86 &0.06   & SC.F & 1  & 27.40 & 0    \\
KC.F & 2  &  24.87 & 0.06   & SI.F &  3 & 26.66 & 0.67   \\
KW.F & 2   & 27.26 &0.06    & SM.F & 2  & 26.85 & 0.21    \\
LB.F & 1  & 24.26 & 0    & TI\_F & 2 & 25.78 & 0.06    \\
 LC.F& 1  & 24.71 & 0    & W.F & 2 & 27.32 & 0.11 \\
LE\_F & 1  & 26.16 & 0  & ZI\_F & 2 & 27.18 & 0.06 \\
LH.F &  2 & 24.68 & 0.17   &     &   &      &    \\

\hline
\end{tabular} 
\caption{Degree, strength, and betweenness of future contracts in the period 1998.09.01 - 2007.12.14. }
\label{tab:centrality}
\end{table}

\section{Time evolution}
Choosing a time window $\Delta T$ for correlations calculation is always a compromise choice between a level of noise from one side and a good estimation of temporal correlation from the other. Increasing $\Delta T$ reduces the noise level, but it gives an average correlation coefficient of the whole window. The correlation coefficients evolve in time and the corresponding MST shrinks during a stock market crisis \cite{Onnela}. Some dynamical aspects of correlations were investigated in ref. \cite{Onnela2,Tumminello}.

\begin{figure}
  \includegraphics[scale=0.35]{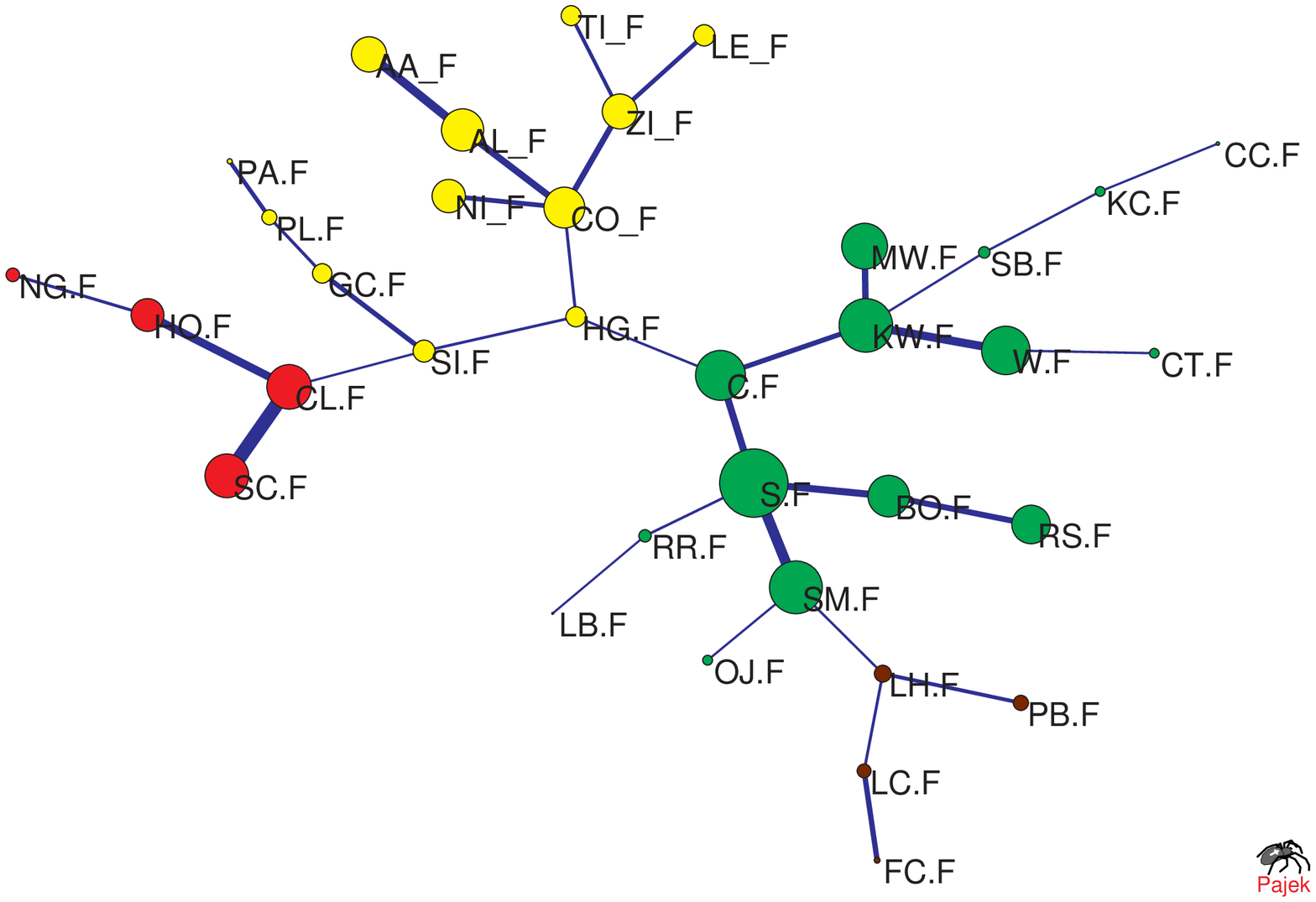}
\caption{(Color online) MST with weights and none strengths for the period 1989.01.03 - 2001.05.10.}
\label{fig:sw1}
\end{figure}

\begin{figure}
  \includegraphics[scale=0.35]{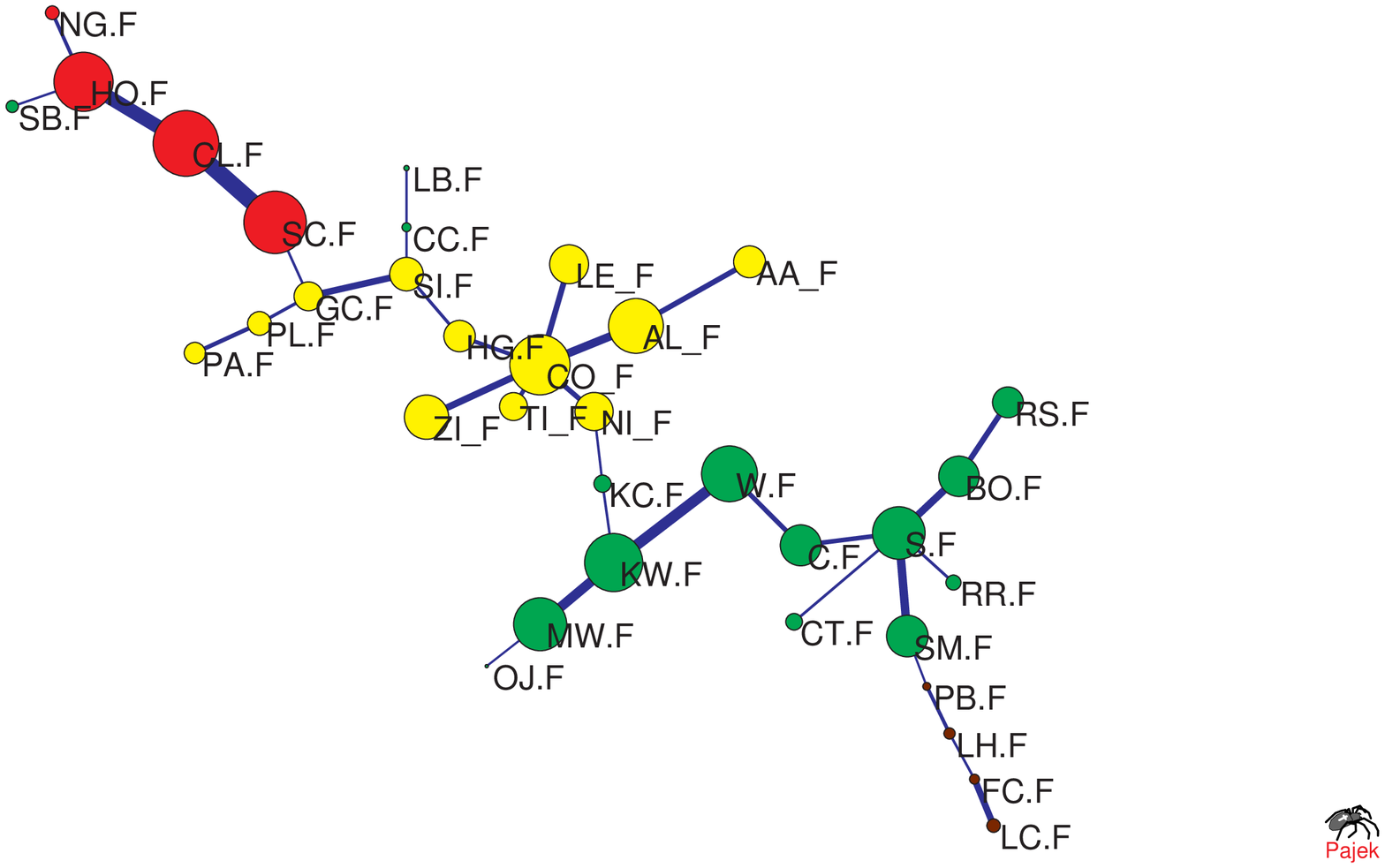}
\caption{(Color online) MST with weights and none strengths for the period 1989.01.03 - 2001.05.10}
\label{fig:sw2}
\end{figure}

\begin{figure}
  \includegraphics[scale=0.3]{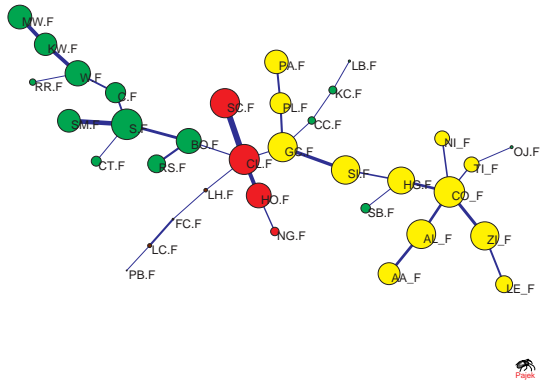}
\caption{(Color online) MST with weights and none strengths for the period 2004.11.08 - 2007.12.14}
\label{fig:sw3}
\end{figure}

We divided the period 1998.09.01 - 2007.12.14 into three subperiods of equal lengths and created MST for this data, computing also strengths of the contracts (fig. \ref{fig:sw1} - \ref{fig:sw3}). The tree changed, but branch clusterization remained.

\begin{figure}
 \includegraphics[scale=0.5,angle=-90]{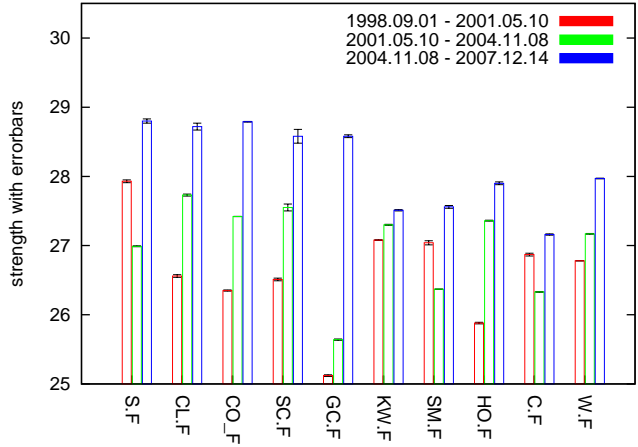}
\caption{(Color online) Change in strength of the strongest contracts with errorbars included. The most noticeable is the increase of gold strength.}
 \label{fig:sily}
\end{figure}

A change in strength of the strongest nodes is presented in fig. \ref{fig:sily}. To check whether the change is caused by a real trend, and not by fluctuations, we recalculated the strength for the same time window starting 7 days earlier, and 7 days later. The results, providing the scope of the strength change, enabled us to include errorbars in figure \ref{fig:sily}. The errorbars appeared to be negligible compared to the change of the strength. It means that while shifting the time window by 7 days the strength remained almost unchanged. All presented contracts increased their strengths, especially gold whose increase was most abrupt. 

\begin{figure}
 \begin{center}
\includegraphics[scale=0.5, angle=-90]{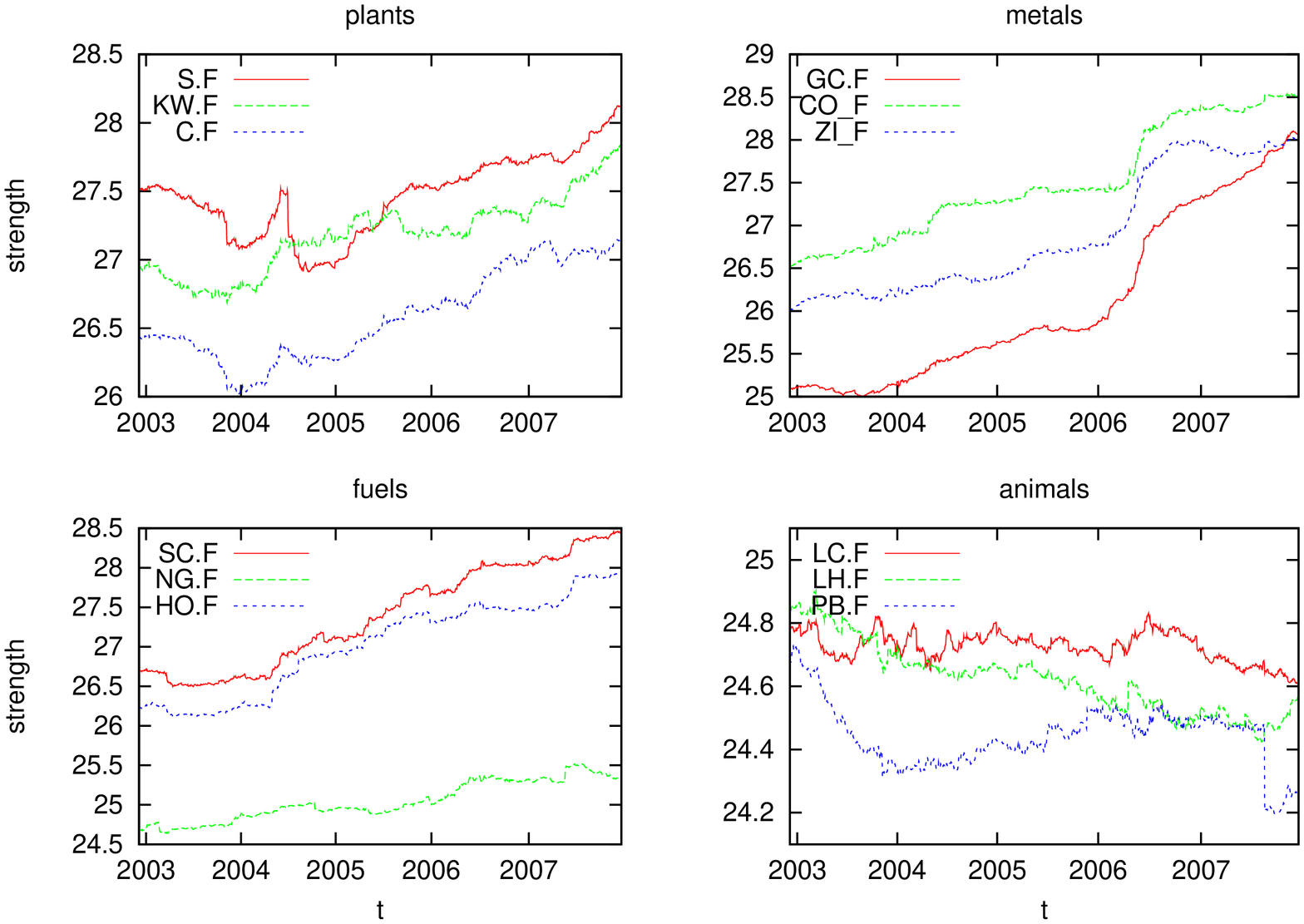}
 \caption{(Color online) Evolution of the strength parameter of selected contracts for a time window $\Delta T=1000$. }
 \label{fig:st_ew}                  \end{center}
\end{figure}

A more detailed view on the strength evolution can be obtained by applying a moving time window analysis. We calculated strengths of contracts for a time window $\Delta T=1000$ (fig. \ref{fig:st_ew}). Contracts of the same branch obey similar changes. Except for animal products, which did not change significantly, all contracts increased their strengths within the investigated period. Also the shape of the evolution curve proves that the observed change is the effect of real trend rather than fluctuations.

\begin{figure}
\begin{center}
\includegraphics[scale=0.5, angle=-90]{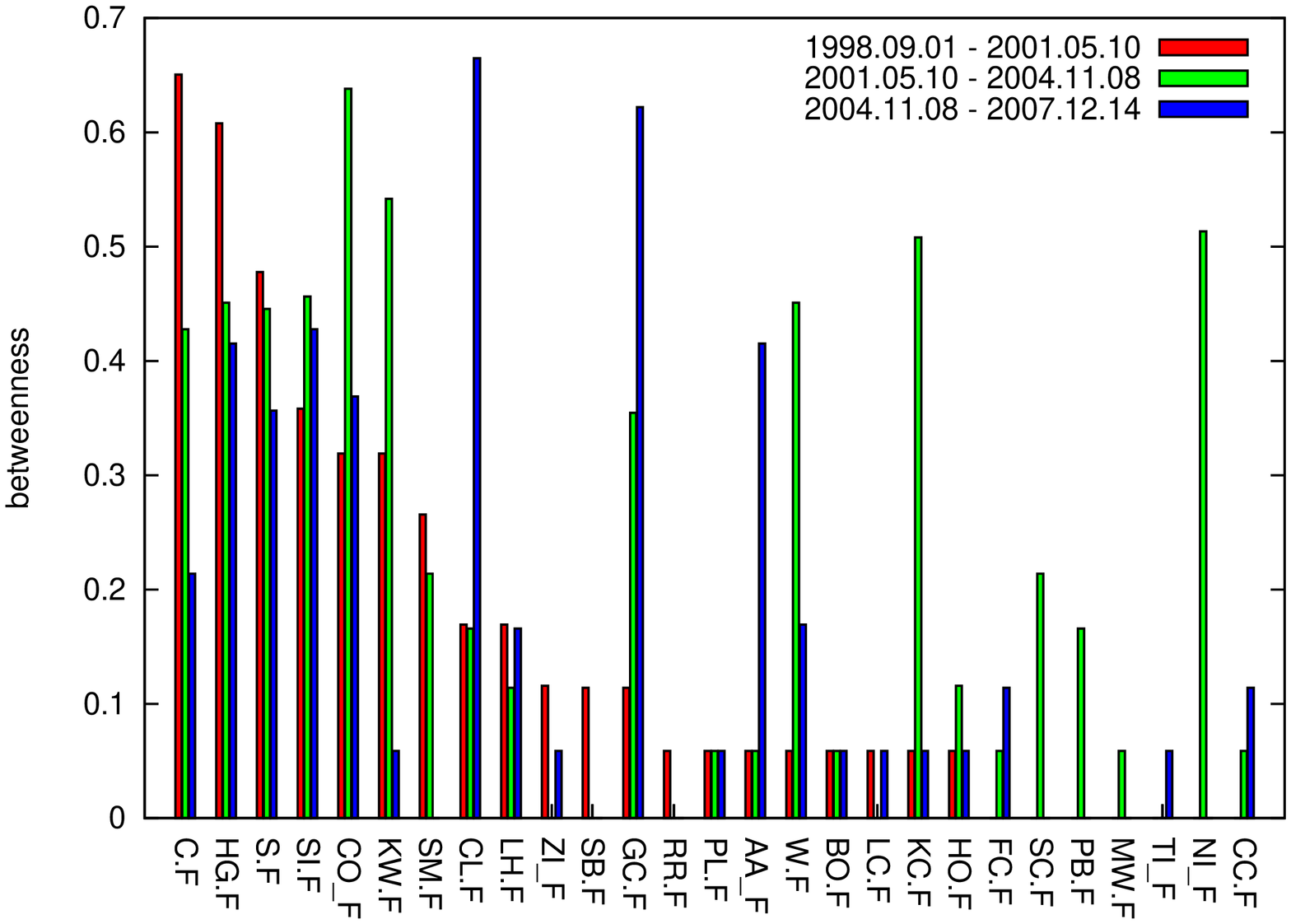} 
\caption{(Color online) Betweenness of future contracts for three subperiods. Contracts that had betweenness $0$ during all three subperiods are skipped in this picture.	}           
\label{rys_betweenness}
\end{center}
\end{figure}

We also calculated betweenness centrality for each tree treated as an unweighted network (fig. \ref{rys_betweenness}). One can observe a monotonic betweenness decrease of corn, copper, soybean, and an increase of gold betweenness. There can also be observed an abrupt increase of crude oil betweenness. 

The change in the picture of correlations  can be caused by noise fluctuation or by a real evolution of market trends. Because our time series are long, compared to their numbers ($T\gg N$), we suspect that we observed a trend evolution rather than fluctuations. 

\begin{figure}
\includegraphics[scale=0.33, angle =-90]{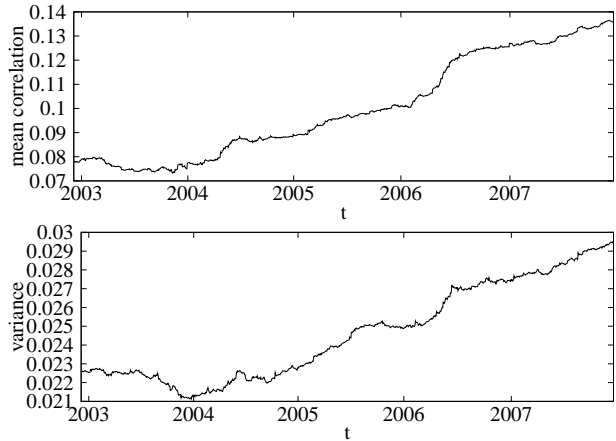} 
\caption{Mean correlation ($\bar C^T(t)$) and variance $\sigma_C$ of $35$ contracts for $\Delta T=1000$ and period 1998.09.01 - 2007.12.14. Because it requires $1000$ records to calculate $\bar C^T(t)$ for a given window $\Delta T=1000$, the figure begins with the end of the year 2002.  On the horizontal axis $t=2003$, $2004$, $...$ means the first (trading) day of a year.}                                                                              
\label{rys_mc}
\end{figure}

We investigated the evolution of the mean correlation,
\begin{equation}
\bar C^T(t)=\frac{2}{N(N-1)}\sum_{i<j}C_{i j}^T(t),
\label{mean_corr}
\end{equation} 
where $C_{i j}^T(t)$ is a correlation coefficient for a time window $[t-\Delta T, t]$. A constant upgoing trend is visible (fig. \ref{rys_mc}). Its interpretation is quite obvious - the mean correlation increased significantly in the investigated period.  We also calculated the variance of the correlation coefficients:
\begin{equation}
\sigma^2_C=\frac{2}{N(N-1)}\sum_{i<j} (C_{ij}^T(t)-\bar C^T(t))^2.
\end{equation}
It is usually positively correlated with the mean correlation \cite{Onnela3}. 

\begin{figure}
\includegraphics[scale=0.33, angle =-90]{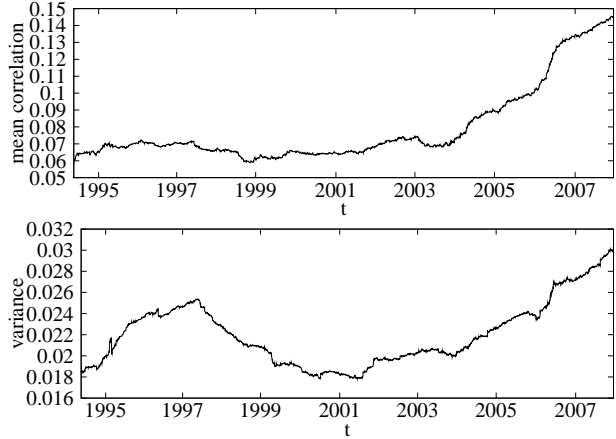} 
\caption{Mean correlation ($\bar C^T(t))$ and variance $\sigma^2_C$ of $27$ contracts for $\Delta T=1000$ and longer period 1990.04.03 - 2007.12.14. }                                                                                                                               
\label{rys_mc2}
\end{figure}

During financial crashes a growth of the mean correlation can be observed \cite{Onnela3}. We can see in the investigated period a constant growth of $\bar C(t)$. We calculated $\bar C^T(t)$ and $\sigma^2_C$ for a longer period, i.e. 1990.04.03 - 2007.12.14 with a smaller set of contracts (fig. \ref{rys_mc2}). We used only $27$ contracts that had been traded through the whole period. The mean correlation was fluctuating between $0.06$ and $0.07$ until the year 2003, next, it started a constant growth to the value $0.14$ at the end of 2007.


\begin{figure}
\includegraphics[scale=0.6, angle=-90]{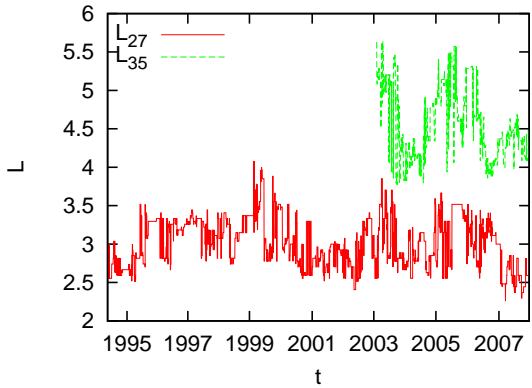} 
\caption{(Color online) The mean occupation layer $L_{35}$ for all 35 contracts, and $L_{27}$ for a reduced sample of contracts in a longer period. The data of $L_{35}$ for better visibility were shifted upwards by an additive constant equal to 1. The values were computed for the time window$\Delta T=1000$.}
\label{fig:ly}
\end{figure}

Let us define the mean occupation layer \cite{Onnela3}:
\begin{equation}
L=\frac{1}{N}\sum_i l(v_i),
\label{mol}
\end{equation}
where $l(v_i)$ denotes the level of vertex $i$. The level measures the distance (in nodes) to the central vertex (which is of level 0). Each time we chose as the central vertex one that minimizes the mean occupation layer $L$. This measure characterizes compactness of MST, which usually shrinks during abrupt price changes (eg. crisis). This effect can be observed as a decrease of the $L$ value. Evolution of mean occupation layer $L$ is presented in fig. \ref{fig:ly}. One can see large fluctuations of both values. No constant trend can be found. Both values are correlated with the correlation coefficient $\rho=0.53$.

\begin{figure}
 \includegraphics[scale=0.5, angle=-90]{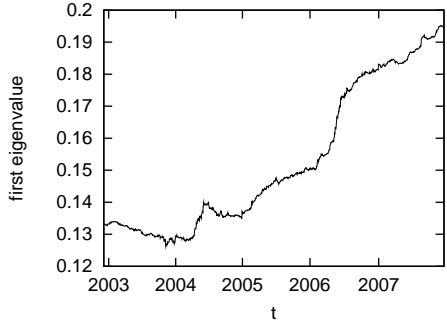}
 \caption{Evolution of the largest eigenvalue normalized by the sum of all eigenvalues for a moving time window $\Delta T=1000$.}
 \label{fig:ew_f}
\end{figure}

The arbitrage pricing theory \cite{Ross, econoph} modells returns of financial assets as a linear function of different macro-economic factors. Such factors can be identified with eigenvectors of a correlation matrix. The largest eigenvector corresponds to an economic factor which has the most significant impact on the market. The level of this impact can be measured by a corresponding eigenvalue normalized by the sum of all eigenvalues. If we take a portfolio of contracts which is a mixture of eigenvectors with equal weights, this value would present the fraction of portfolio variance explained by the first factor. The evolution of normalized largest eigenvalue is presented in fig. \ref{fig:ew_f}. The observed increase means that the influence of the first factor became even stronger with time, which caused the observed increase of the mean correlation (fig. \ref{rys_mc}).

\begin{figure}
 \begin{center}
\includegraphics[scale=0.5, angle=-90]{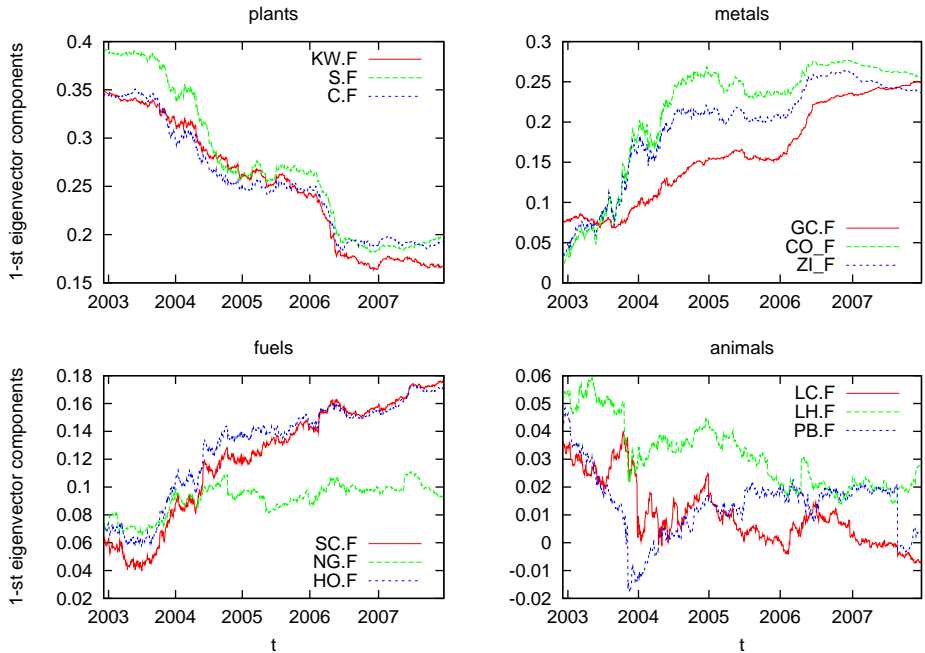}
 \caption{(Color online) Contribution of selected contracts to the eigenvector corresponding to the largest eigenvalue which is normalized to unity ($\Delta T=1000$).}  
\label{fig:ew_f2}                                                                                                                      \end{center}
 
\end{figure}

We plotted evolution of the first eigenvector components (fig. \ref{fig:ew_f2}) which can be treated as an influence of the first factor on a given commodity. It strongly decreased for plant products, increased for metals and fuels, and remained almost unchanged for animals. While growing influence of the first factor resulted in the increase of mean correlation, changes of eigenvector components drove the evolution of strength parameters. It seems to be clear for metals, fuels and animals. However, for the case of plants we observed a decrease of first eigenvector components. At the same time the largest eigenvalue increased. The two opposite tendencies resulted in slight increase of strengths of plants. 

\section{Discussion and conclusions}

 We analyzed the correlation matrix of commodity prices. Due to long time series the level of noise in correlation matrix was low. Two eigenvectors of $C_{ij}$, corresponding to the largest eigenvalues were plotted. The picture (\ref{rys_eigenvectors}) visualizes a clustering structure of correlations. Using a correlation metric $d_{ij}$ we created MST of investigated contracts. MST provides clear evidence for the existence of strong correlations of commodities within a given sector and for intersector correlations depending on the level of similarity. 

The above picture of correlations could be expected from similar studies of stock markets and currencies. More interesting are investigated dynamical properties of correlations which, as we observed, changed in time. In the studied period i.e. 1998.09.01 - 2007.12.14 the mean correlation (\ref{mean_corr}) increased from about $0.08$ to about $0.13$. The reason for this correlation increase can be as follows. Commodity markets attracted a lot of investors' attention in the last years. A growing demand for energy, metals, and food from fast developing Chinese and Indian economy created a boom for commodities and an impulse for intensive speculations. This market boom makes prices follow one direction, which resulted in increase of correlation between the traded assets. A similar effect was observed during market crashes \cite{Onnela}. The effect of the boom for commodities is responsible for the observed growth of mean correlation coefficient. 

The mean correlation calculated as an average measure increased, but the contributions of individual contracts to this difference were various. We introduced strength of a contract expressing the magnitude of its correlation with other contracts. Strengths calculation for moving time window visualised the evolution of this parameter. It turned out that metals and fuels importantly participated in the increase of strength, while animals almost remained unchanged. 

It is obvious that prices of financial assets are driven by several economic factors. Among them there is the most significant factor, called by us the first factor. A measure of importance of the factor can be a corresponding eigenvalue. We showed that the role of the first factor was getting stronger, so the market finally became more and more one-factored. This fact, that can be observed in the increase of the first eigenvalue, is also responsible for the increase of correlations, characterized by the mean correlation parameter. That effect was caused by a growing price bubble.

 We also observed the change of the first factor itself, or from the other viewpoint, the change of its influence on specific commodities. The first eigenvector components evolved in time in a way that depends on the commodity branch. Contracts which became significantly more dependend on the first factor, that is metals and fuels, became also stronger. Contracts with weak interactions with the first factor, like animals, did not change their relatively low strengths.  

We studied correlations of commodity contracts returns. Our analysis showed that starting from the year 2003 the commodity market became more correlated, and was driven by a single economic factor. The dependence on the factor got stronger with time. The result is most pronounced in an evolution of the largest eigenvalue normalized by the sum of all eigenvalues, and it can be also seen in the evolution of contracts strengths and their mean correlation.

The most important result seems to be the clear evidence of the constantly increasing market synchronization. This fact can be even related to the global scale of current economic crisis. However, the authors cannot suggest that they found roots of recent dramatic market changes. 

It was expected that dynamics of commodities belonging to the same sector would be clustered, i.e. would be closely placed in the MST. One could however not predict in advance which of commodities would become synchronized with the first eigenvalue (a market mode). It is surprising that the agriculture products behaved differently compared to metals that became much more synchronized.

An obvious implication from our study is the increasing risk for market players that cannot effectively diversify their portfolios, since corresponding assets are more and more correlated. It is well known that such correlations modify substantially the optimal portfolios and make diversification more difficult since the effective number of asses became smaller \cite{Bouchaud}.

\section*{Acknowledgements}
We acknowledge a support from the EU Grant Measuring and Modeling
Complex Networks Across Domains — MMCOMNET (Grant No. FP6-2003-
NEST-Path-012999) and from Polish Ministry of Science and Higher Education (Grant No. 13/6. PR UE/2005/7).

\end{document}